\documentclass[paper,11pt]{JHEP}
\usepackage[centertags]{amsmath}
\usepackage{amsfonts} 
\usepackage{graphicx}

\def\one{{\rm 1\kern -.9mm l}}                             %

\def\beq{\begin{equation}}
\def\eeq{\end{equation}}
\def\beqa{\begin{eqnarray}}
\def\eeqa{\end{eqnarray}}
\def\um{\frac12}
\def\p{\partial}
\newcommand{\eqa}{\begin{eqnarray}}
\newcommand{\ena}{\end{eqnarray}}

\newcommand{\Tr}{\mathrm{Tr}\,}

\newcommand{\cL}{\mathcal{L}}

\newcommand{\mc}[1]{\mathcal{#1}}
\newcommand{\tr}{\textup{Tr}}

\def\ep{\epsilon}
\newcommand{\de}{\delta}
\newcommand{\pa}{\partial}

\title{Lorentz completion of effective string\\ (and p-brane) action 
}
\author{ Ferdinando Gliozzi and Marco Meineri
\\
\vskip 0.2cm
Dipartimento di Fisica Teorica, Universit\`a di Torino\\
and Istituto Nazionale di Fisica Nucleare - sezione di Torino \\
Via P. Giuria 1, I-10125 Torino, Italy\\
\vspace{0.25cm}
\email{gliozzi@to.infn.it marco.meineri@studenti.unito.it} 

}
\abstract{ The formation of a confining string (or a p-brane) in a Poincar\'e
invariant theory breaks spontaneously this symmetry which is thereby realized non-linearly in the effective action of these extended objects. 
As a consequence the form of the action is strongly constrained. A new general method is described to  obtain 
in a systematic way higher order Lorentz invariant contributions to 
this action. We find a simple recipe to promote a term invariant under the stability subgroup to an expression invariant under the whole Lorentz group. It is 
based on the following three steps: in the saturation of worldsheet (or worldvolume) indices replace the Minkowski metric with the inverse of the induced metric;  in the saturation of  indices of the transverse coordinates describing the position of the extended object replace the 
Euclidean metric with a certain new metric; finally replace the field derivatives of order higher than two with a certain covariant 
derivative. Lorentz invariance of the expression modified this way 
immediately follows. We find in particular that the leading bulk deviation 
of the Nambu Goto action in any space-time dimensions is proportional to the square of scalar curvature.  
}
\keywords{Bosonic Strings, Nonlinear Lorentz realization, Effective actions}
\begin{document}
\section{Introduction}
\label{sec:introduction}
The effective string action describes one-dimensional solitonic objects embedded in a space-time of higher dimensions, like for instance 
Abrikosov vortices in superconductors, Nielsen Olesen vortices 
in Abelian Higgs theory or cosmic strings in grand unified models. It is particularly useful in the study of the long-distance properties of the string-like flux tubes in confining gauge theories \cite{Luscher:1980fr}  where, even though the formation of such an extended object between quark sources is not a 
proved fact, numerical experiments and theoretical arguments in lattice gauge theories leave little doubt that this physical picture is basically correct 
\cite{{Caselle:1996ii},{Luscher:2002qv}, {Luscher:2004ib},{Teper:2009uf},{Pepe:2010na}}.

When this string-like object forms in the vacuum, the Poincar\'e symmetry breaks spontaneously and  Nambu-Goldstone modes appear. These are described, in a $D$ -dimensional Minkowski space-time,  by the transverse displacements 
$X_i(\xi_0,\xi_1)$ of the string $(i=2,\dots,D-1)$. Integrating 
out the massive degrees of freedom of the $D$-dimensional theory one ends up, at least in principle, with a two-dimensional effective action 
\beq
S=\sigma\int d^2 \xi \,\cL(\partial_aX_i,\partial_b\partial_cX_j,\dots)\,,
\label{effaction}
\eeq      
where $\sigma$ is the string tension and the Lagrangian density is a local function of the derivatives of the transverse fields $X_i$. It is expedient to think 
of $S$ as the  low energy effective action describing the fluctuations of a 
long string of length $R$. Then there is a natural dimensionless expansion parameter \cite{Caselle:1994df}, namely $1/(\sqrt{\sigma}\,R)$, and this expansion 
corresponds to the power expansion of $S$ in the number of derivatives.

The physical behaviour of the string cannot depend on the choice of its worldsheet coordinates, of course. In fact  this theory could be written in a form invariant under diffeomorphisms. The present formulation, called physical or static gauge, uses a choice of coordinates  in which only the physical degrees of  freedom - the transverse fluctuations $X_i$ of the string- appear in the action. 

Notice that the process  of dimensional reduction from a $D-$ dimensional fundamental theory to an effective two-dimensional string action is not always at a purely conjectural stage. In the physics of interfaces in three dimensional systems, for instance, in some cases one is able  to integrate out bulk degrees of freedom \cite{{Provero:1995cz},{Jaimungal:1998hk}}, obtaining in this way the Nambu-Goto action \cite{Jaimungal:1998hk} or its Gaussian limit \cite{Provero:1995cz}.

The terms contributing to the effective action are all those respecting 
the symmetries of the system, thus the worldsheet derivatives $\partial_a$ 
$(a=0,1)$  should be saturated according to the worldsheet symmetry $SO(1,1)$ and similarly the transverse indices $i=2,\dots,D-1$ should always form scalar 
products in order to respect the transverse $SO(D-2)$ invariance. Besides 
these obvious symmetries there are some further constraints to be taken 
into account.
As first observed in \cite{Luscher:2004ib}  and then generalized in 
\cite{Aharony:2009gg}, comparison of the string partition function in different channels (``open-closed string duality'') fixes by consistency  the first 
coefficients, at least, of the derivative expansion of $S$. It was subsequently recognised the crucial role of the Lorentz symmetry of the underlying Yang-Mills  theory \cite{{Meyer:2006qx}, {aks},{Aharony:2010cx}}. In fact, integrating out the massive modes of the fundamental theory to obtain the effective string action  does not spoil its $ISO(1,D-1)$ symmetry, but simply realises it in a non-linear way, as always it happens in the spontaneous breaking of a continuous symmetry \cite{{Coleman:1969sm},{Callan:1969sn}}. In other words, the physical gauge hides the Lorentz invariance of the underlying theory which is no 
longer manifest, but poses strong constraints on the form of the effective 
action, as  this should be invariant under the effect of a non-linear Lorentz transformation. The latter, when applied to a generic term made with $m$ derivatives and $n$ fields, schematically $d^m X^n$, transforms it in other terms with the same value of the difference 
$m-n$, called ``scaling'' of the given term \cite{Aharony:2011gb}.
The terms of scaling zero are only made  with  first derivatives  and demanding invariance under the non-linear Lorentz transformations implies that they coincide  with the derivative expansion of the Nambu-Goto action \cite{{aks},{Aharony:2010cx},{Gliozzi:2011hj}}. There is however a subtlety about this point that should be mentioned. The first few terms of the derivative expansion of the effective action (\ref{effaction}) associated with a  Wilson loop encircling a 
 minimal surface $\Sigma$ of area  $A$ are, omitting the perimeter term,
\beq
S=-\sigma A-\int_\Sigma d^2\xi\left(\frac{c_0}2 \p_aX\cdot\p^aX+
c_2(\p_aX\cdot\p^aX)^2+c_3(\p_aX\cdot\p_bX) (\p^aX\cdot\p^bX)+\dots \right)
\eeq
where  $c_i$ are dimensionful parameters. Note that the $X_i$'s have the dimensions of length, as they are the transverse displacements of the string, hence $c_0$ cannot be reabsorbed in a redefinition of $X_i$ and gives measurable 
effects. In particular the transverse area $w^2$ of the flux tube increases logarithmically
 with the quark distance $R$ and at the leading order one finds \cite {{Luscher:1980iy},{Luscher:1980ac}}
\beq
w^2=\frac{D-2}{2\pi c_0}\log(R/R_0)\,,
\eeq
where $R_0$ is a low-energy distance scale. In the Nambu-Goto string it 
turns out that $c_0=\sigma$ and this has been confirmed by 
numerical calculations in different lattice gauge models \cite{{Caselle:1995fh},
{Giudice:2006hw},{Gliozzi:2010zv}, {Gliozzi:2010zt}}. The last equality 
 is not a specific property of the Nambu-Goto model: open-closed string 
duality implies $c_2=c_0/8$, $c_3=-c_0/4$  and
$c_0=\sigma$ as is easy to verify. On the contrary, at the classical level, 
the only requirement of Lorentz invariance, even if it fixes the whole 
series of scaling zero terms, does not link $c_0$ to the string tension. 
We find indeed (see next Section)
\beq
S=-c_0\int_\Sigma d^2\xi\sqrt{-g}+(c_0-\sigma)\int_\Sigma d^2\xi+
{\rm higher ~scaling~ terms}\,,
\label{zeros}
\eeq 
where $g$ is the determinant of the induced metric defined in (\ref{indug}).
Clearly (\ref{zeros}) can be put in a reparametrization invariant form only if 
$c_0=\sigma$.

In order to find the explicit form of Lorentz invariant higher order terms
in (\ref{zeros}) one starts typically with 
a non-vanishing term of scaling greater than zero and adds iteratively an infinite sequence of terms generated by the 
non-linear Lorentz transformation. Only if this process comes to an end and 
no further terms are generated 
one can conclude that the starting term has a Lorentz-invariant 
completion and is then compatible 
with Lorentz symmetry. This procedure has been accomplished 
for the boundary action of the open string 
in \cite{Billo:2012da}, where a systematic classification of the Lorentz invariant contributions up to terms of scaling six has been found. 

So far, the form of the leading bulk correction to the Nambu-Goto action is still debated. A class of
obvious higher order Lorentz invariants can be constructed in terms of the extrinsic curvature 
\cite{{Polyakov:1986cs},{Kleinert:1986bk}}
 or other geometric quantities like (powers of)  Gaussian curvature of 
the induced metric \cite{Gomis:2012ki}, but these geometric terms do not 
exhaust the list of the Lorentz invariants, as we shall show 
in the present paper.
So far it has been assumed that the first allowed correction to the Nambu-Goto action 
could be the six derivative term \cite{Aharony:2009gg}
\beq
S_4=-c_4\int d^2\xi\left(\p_a\p_b X\cdot\p^a\p^b X\right) (\p_cX\cdot\p^cX)\,,
\label{s4}
\eeq
 which is non-trivial only when $D>3$. Opinions differ on the role
of this term and on the value of the coefficient $c_4$. In \cite{Aharony:2011gb} it was suggested 
that if this term has an all-orders Lorentz invariant completion, it could be the analogous in the 
static gauge of the  contribution conjectured by Polchinski and Strominger
 \cite{Polchinski:1991ax} as the leading correction to the Nambu- Goto action in the conformal gauge,
where the Lorentz symmetry is linearly realized.
In this gauge it takes the form
\beq
\frac{26-D}{96\pi}\int d^2\xi\sqrt{-g}R\frac1{\Box}R\,,
\eeq
where $R$ is the induced curvature scalar and $\Box$ the d'Alembertian. 
Motivated by this result it was conjectured that $c_4=(26-D)/192\pi$, 
see also \cite{Aharony:2011ga}. In \cite{Dubovsky:2012sh} Eq. (\ref{s4}) is 
instead considered as a  Lorentz-violating counterterm necessary to cancel a Lorentz anomaly 
generated by the $\zeta$-function regularization in the static gauge and the coefficient $c_4$ turns out 
to be $c_4=-1/8\pi$.    

In the present paper  we show in particular  that the above term is 
actually absent, at least at the classical level.
Our goal is to describe a  general class of Lorentz invariants 
which are obtained by  performing an all orders Lorentz completion of suitable 
 terms.
Applying this method to the term (\ref{s4}) we find that the orbit generated by the  non-linear Lorentz transformation includes other terms of the same 
perturbative order which combine with (\ref{s4}) to form a total derivative.
 Thus we are led to conclude that there is really no  six derivative term compatible with Lorentz completion.

The recipe we find to build a Lorentz invariant in the bulk space-time is not specific for the string, as it can be applied to any $p-$dimensional 
classically flat extended object, like for instance  $p-$branes. As expected, in this context the induced metric $g_{ab}$ plays a crucial role. We have
\beq
g_{ab}=\eta_{ab}+\delta^{ij}\p_aX_i\p_bX_j=\eta_{ab}+h_{ab}\,,
\label{indug}
\eeq 
where $\eta_{ab}$ is the diagonal Minkowski metric with
$1=-\eta_{00}=\eta_{aa}$ $(a=1,\dots,p)$ and its matrix inverse is 
$\eta^{ab}$ with $\eta_{ab}\eta^{bc}=\delta^c_a$.
Our recipe  differs from the one suggested by the classical works on non-linear realization of spontaneously broken symmetries 
\cite{ {Coleman:1969sm}, {Callan:1969sn},{Isham:1971dv}}, based on the introduction of suitable covariant derivatives. 

A generic term invariant under the unbroken subgroup $ISO(1,p)\times ISO(D-p-1)$ is formed by scalar products of the worldvolume indices saturated by $\eta^{ab}$ and scalar products 
in the transverse coordinates 
saturated with $\delta^{ij}$. In terms of these quantities the recipe to obtain
a Lorentz invariant is particularly simple if we begin with a seed term in 
which every transverse field $X_i$ appears with two derivatives, at least. 
Then the following two moves are necessary to generate a Lorentz invariant:
\begin{itemize}
\item[$i)$]
 replace in each scalar product of the worldvolume indices 
$\eta^{ab}$ with $g^{ab}$, where
\beq
g^{ab}=\eta^{ab}-\eta^{ac}h_{cd}\eta^{db}+\eta^{ac}h_{cd}\eta^{de}h_{ef}\eta^{fb}-\dots
\eeq
is the matrix inverse of $g_{ab}$; 
\item[$ii)$] replace in each scalar product of the transverse coordinates $\delta^{ij}$ with $t^{ij}$, where 
\beq
t^{ij}=\delta^{ij}-\p_aX^ig^{ab}\p_bX^j\,.
\eeq
\end{itemize}
Clearly the first move corresponds to a resummation of an infinite tower of terms. If the seed term contains also transverse coordinates with more than two derivatives there is a third rule to be added which allows to  lower the order of 
the derivatives. It turns out that these three rules are sufficient to yield
 Lorentz invariants formed by simple combinations of these resummed quantities multiplied by $\sqrt{-g}$.

 Notice that, at variance with what is generally made in this 
context, we  use neither field redefinitions nor  equations of motion or 
integrations by parts, thus  in the derivative 
expansion  of our Lorentz invariants the first few (sometimes all) terms 
may vanish on shell, at least at the perturbative level.  
For instance, at scaling two we find two Lorentz invariants obtained 
by applying the above rules to
the two terms  $\partial_a\partial_bX^k\partial^a\p^bX_k$ and 
$\Box X^k\Box X_k$, which are both vanishing on shell. The two moves $i)$
and $ii)$ yield the following two invariants

\beq
I_1=\sqrt{-g}\Big(\partial_{ab}^2X^k\partial_{cd}^2X_k\, g^{ac}g^{bd}
-\partial_{ab}^2X_k\partial_{cd}^2X_i\partial_eX^k
\p_fX^i\, g^{ac}g^{bd}g^{ef}\Big)\,,
\label{i1}
\eeq
\beq
I_2=\sqrt{-g}\Big(\partial_{ab}^2X^k\partial_{cd}^2X_k g^{ab}g^{cd}
-\partial_{ab}^2X_k\partial_{cd}^2X_i\partial_eX^k
\p_fX^i\, g^{ab}g^{cd}g^{ef}\Big)\,.
\label{i2}
\eeq
In particular we recover the Hilbert-Einstein Lagrangian
\beq
I_1-I_2=\sqrt{-g} R\,,
\label{HE}
\eeq
where $R$ is the scalar curvature of the induced metric
\beq
\begin{split}
R=&(\p_{ab}^2X\cdot\p_{cd}^2X)(g^{ac}g^{bd}-g^{ab}g^{cd})\\
&-(\p_{ab}^2X\cdot\p_eX)(\p_{cd}^2X\cdot\p_fX)g^{ef}
(g^{ac}g^{bd}-g^{ab}g^{cd})\,.
\end{split}
\eeq

Although in the case of the effective string the Hilbert-Einstein Lagrangian 
is a total derivative (there are no handles in the present description of the 
worldsheet), it is no so for a generic $p-$brane. This observation 
illustrates  the fact that our recipe can generate non-vanishing invariants starting from terms which are zero on shell.

In the case of the $p-$branes we can assume that the massless modes 
propagating through the extended object are not only the Goldstone modes, i.e. the transverse coordinates $X_i$ $(i=p+1,p+2,\dots D-1)$ but also a 
$p-$dimensional abelian gauge field $A_a$ $(a=0,1,\dots p)$.  It is easy to 
extend our Lorentz invariants to this more general case by exploiting the fact 
that the way of transforming of the field strength $F_{ab}=\p_aA_b-\p_bA_a$ under a non-linear Lorentz transformation is exactly the same as the induced metric 
$g_{ab}$ \cite{{Gliozzi:2011hj},{Casalbuoni:2011fq},{Asakawa:2012px}}, hence it  suffices 
to replace in our Lorentz invariants the induced metric $g_{ab}$ and/or 
$g^{ab}$ with the combination $e_{ab}=g_{ab}+\lambda 
F_{ab}$ and/or its inverse $e^{ab}$, defined by $e^{ac}e_{cb}=\delta^a_b$, 
to obtain these more general invariants.

The outline of the paper is as follows. In the next section we describe a new way to deal with the scaling zero invariant and introduce a useful diagrammatic representation. In section 3 we derive the first two rules which are necessary to obtain Lorentz invariant expressions and describe the scaling two invariants obtained with this new method. In section 4 we discuss higher scaling invariants,
describe the third rule and apply it to write explicitly  a set of invariants of scaling four. Finally in the last section we draw some conclusions.
\section{Scaling zero }                                                  
It is convenient to introduce a diagrammatic representation of the terms contributing to derivative expansion of the effective action. Each term is 
associated to a graph where the nodes represent the fields $X_i$, and there are two kind of links connecting the nodes. They represent the two types of saturation; solid lines represent saturation of worldvolume indices while wavy lines are associated to the saturation of transverse indices. 

\begin{figure}[html]
\begin{center}
\includegraphics[width=0.5\columnwidth,height=0.1\textheight]{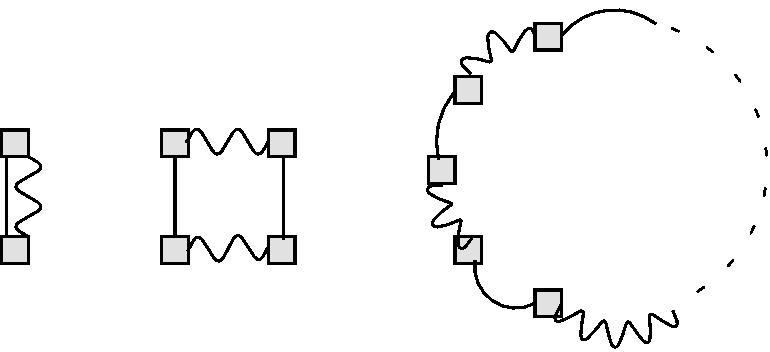}
\end{center}
\caption{Possible structures with first derivatives of the fields. 
Solid lines stand for worldvolume indices, wavy lines for scalar products in the bulk. The generic term at scaling zero is a product of rings of 
different sizes.}
\label{fig:rings}
\end{figure}

     At scaling zero there is just one possible structure. A graph at any order in the derivative expansion is a product of \emph{rings}, i.e. polygons with an even number of vertices, while the links alternate solid and wavy lines as shown in fig.\ref{fig:rings}.  

\begin{figure}[ht]
\begin{center}
\includegraphics[width=0.8\columnwidth,height=0.2\textheight]{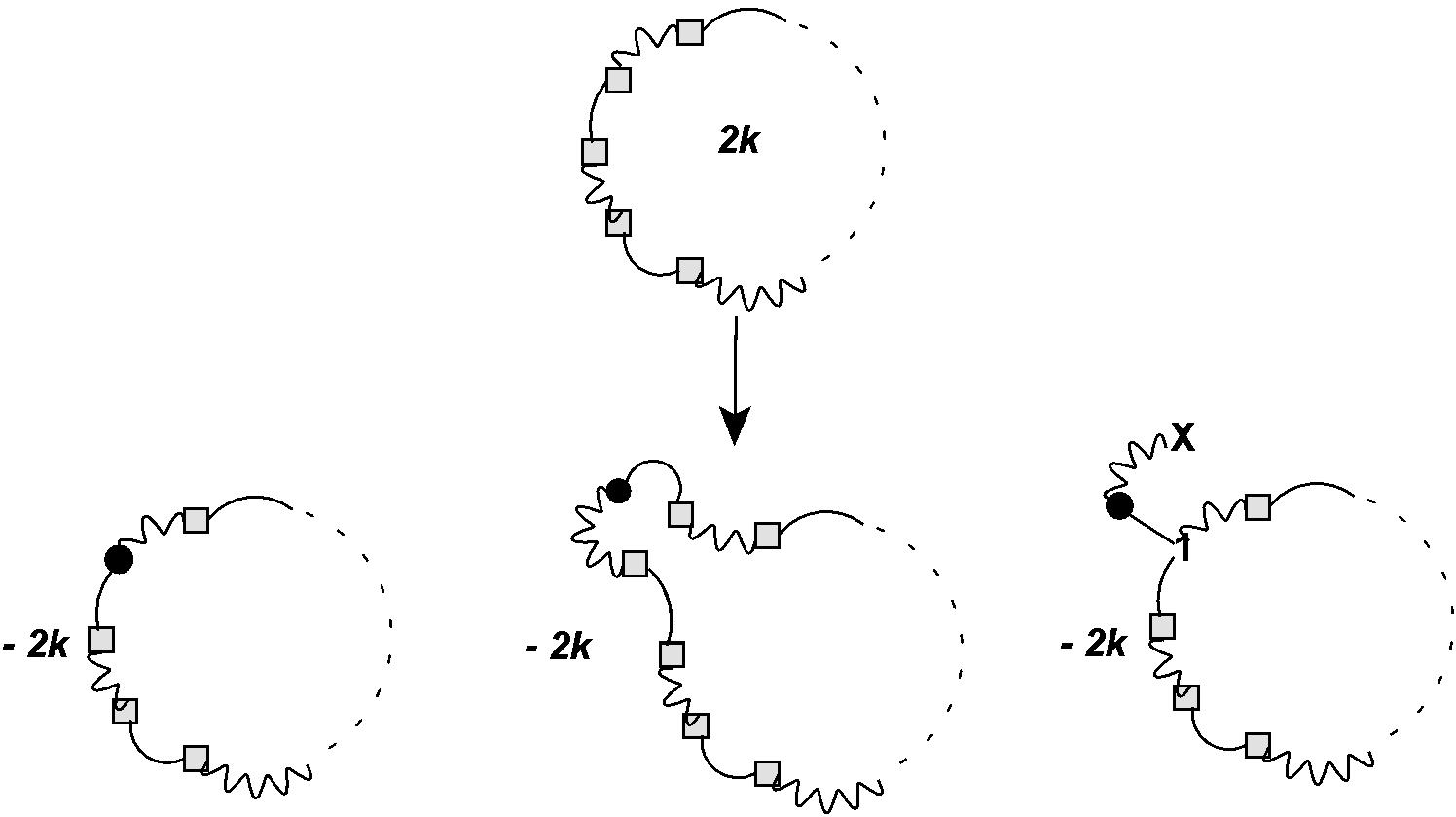}
\end{center}
\caption{Variation of a ring at order $2k.$ The dot represents the 
matrix $\ep^{aj}$ of parameters of the transformation.}
\label{fig:varring}
\end{figure}

\begin{figure}[ht]
\begin{center}
\includegraphics[width=0.8\columnwidth,height=0.2\textheight]{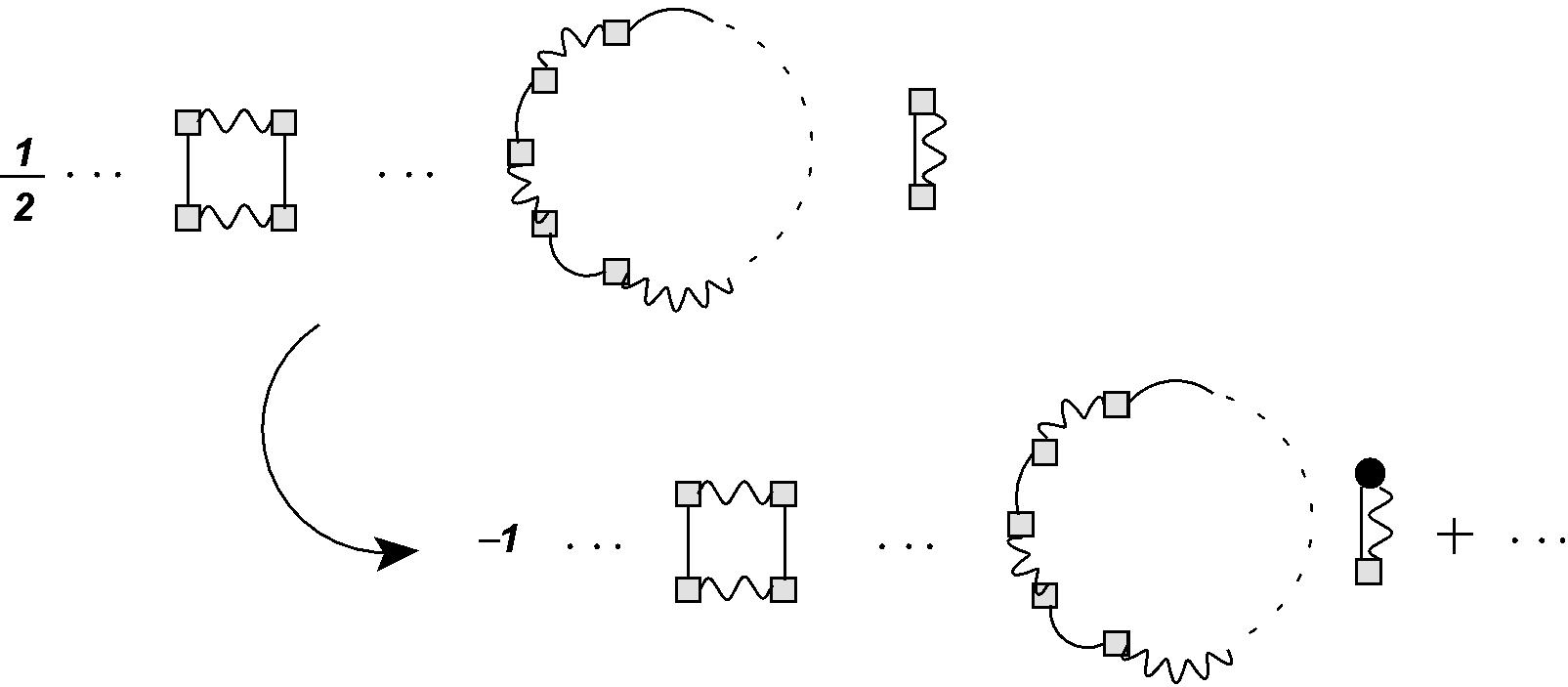}
\end{center}
\caption{Graph required to sweep out the $X_i$ contribution coming from a term containing neither the last 
$\pa_aX\cdot\pa^aX$ factor nor the $\frac12$ coefficient.}
\label{fig:newring}
\end{figure}

It is useful to write the non-linear infinitesimal Lorentz transformation
 \cite{Aharony:2010cx} in a covariant form 
\beq
\de X^i=-\ep^{aj}\de^{ij}\xi_a-\ep^{aj}X^j\pa_aX^i\,,
\eeq
so that
\beq
\de \left(\pa_bX^i\right)=-\ep^{aj}\de^{ij}\eta_{ab}-\ep^{aj}\pa_bX^j\pa_aX^i-\ep^{aj}X^j\pa_a\pa_bX^i\,.
\label{variation}
\eeq
As a consequence, a variation of a ring is shown in fig.\ref{fig:varring}. 
The first two addends provide the recurrence relation. They must cancel 
order by order independently from other terms that could be multiplied 
with the ring under consideration. Therefore, if we could neglect the 
third addend, we would sum up the series associated with a ring, 
and this would be enough. The third variation, on the contrary, 
must be cancelled adding the missing terms to form a total derivative. 
We get a total derivative by moving the solid link of the dot around 
from one vertex to the other. In the case of
 a product of rings, the variation of every ring provides the right 
contribution, and the total derivative is found adding one more graph 
(see fig.\ref{fig:newring}). So this request forces us to add a 
new minimum ring, and with it  the whole tower of growing rings to cancel 
the first two addends in fig.\ref{fig:varring}. However, 
the situation is kept simple by the fact that linearity allows us to 
sum up the series associated with a ring, and then add 
to the result the new ring to form the total derivative. 

Let us concentrate on a single ring. Referring again 
to fig.\ref{fig:varring}, the compensation of the first 
two addends order by order leads to the recurrence relation
\beq
(k+1)a_{k+1}=-k\,a_k,
\eeq
where the coefficient $a_k$ is associated with the ring with $2k$ vertices. 
The solution is
\beq
a_k={(-1)}^{k+1}\frac{1}{k}\,a_1.
\label{logtayl}
\eeq
So we get the series
\beq
\begin{split}
a_1&\sum_{k=1}^\infty {(-1)}^{k+1}\frac{1}{k} \big[{(\pa X\cdot\pa X)}^k\big]^a_{\ b}\, \de^b_{\ a}=
a_1\tr\left[\log\left(\mathbb{1}+\pa X\cdot\pa X\right)^a_b\right] = \\
&= a_1\log\bigr\{\det\left({\mathbb{1}}+\pa X\cdot\pa X\right)^a_{\ b}\bigr\} = 
a_1\log\bigr\{\det\left[\eta^{ac}\left(\eta+\pa X\cdot\pa X\right)_{cb}\right]\bigr\} = \\
&a_1\log\bigr[-\det\left(\eta_{ab}+h_{ab}\right)\bigr]=\log(-g)\,,
\end{split}
\eeq
 where we have used the induced metric $g_{ab}$ defined in (\ref{indug}). 
To deal with total derivatives, it is sufficient to consider a sequence 
$b_n$ whose index counts the number of logarithms. The general 
addend of the series is then $b_n\bigr[\log(-g) \bigr]^n$.

Now consider the derivative expansion of the logarithms again. 
At order $n$ there are $n$ rings, and if we choose a size for each of them, 
the corresponding graph appears in the expansion $n!$ times. 
To build a total derivative we must add a graph with the same $n$ rings, 
plus one ring with just two vertices. This new graph comes from
 the order $b_{n+1}$ with a multiplicity factor $(n+1)!$. There is no 
other numerical factor from the Taylor series of the logarithms to 
deal with, as we normalised all $a_1$'s of \eqref{logtayl} to one. 
Taking into account the $\frac12$ factor from fig.\ref{fig:newring}, 
we find the recursion relation
\beq
b_{n+1}=\frac{1}{2(n+1)}b_n,
\eeq
whose solution is
\beq
b_n=\frac{1}{2^n\,n!}\,b_0.
\eeq
Now we can write the Lagrangian density at scaling zero
\beq
\mc{L}_0=\ b_0\sum_{n=0}^\infty\frac{1}{n!}\left[\frac{1}{2}
\log(-g) \right]^n-b_0  =b_0\sqrt{-g}-b_0\,.
\eeq
Choosing $c_0=-b_0$ we recover Equation (\ref{zeros}) quoted in the Introduction. 

An apparent weak point in the above calculation is the implicit 
assumption that all the rings are algebraically independent, while in 
a $p-$brane only the first 
$p+1$ are so. It would be not difficult to fix  this point in the 
general case \cite{Meineri}, however we believe that it is more 
useful and instructive to concentrate on the case of the effective string 
where we can give a complete, alternative proof. In the latter case there are only two 
independent ring terms, namely $\tr h=h^a_a$ and $\tr (h^2)=h^a_bh^b_a$. 
 For the other rings it is not difficult to verify that
\beq
\tr h^n=\left(\frac{\tr h+\sqrt{2\tr( h^2)-(\tr h)^2}}2\right)^n+
\left(\frac{\tr h-\sqrt{2\tr (h^2)-(\tr h)^2}}2\right)^n\,,
\eeq  
thus the most general scaling zero expression is
\beq
I_0=\sum_{m=0}^\infty\sum_{n=1}^\infty c_{n,m}(\tr h)^n[\tr (h^2)]^m=
\sum_{m=0}^\infty\sum_{n=0}^\infty c_{n,m}(\tr h)^n[\tr (h^2)]^m-c_{0,0}\,,
\eeq
where we added and subtracted the constant term $c_{0,0}$ in order to simplify the solution of the recursion relations dictated by  Lorentz invariance. We find
\beq
(n+1)c_{n+1,m}+\left(\frac n2+m-\frac12\right)c_{n,m}+(m+1)c_{n-1,m+1}=0\,,
\eeq
and
\beq
(n+2)c_{n+2,m}+\left(m-\frac12\right)c_{n+1,m}-(m+1)c_{n-1,m+1}=0\,,
\eeq
with the initial conditions
\beq
c_{n,m}=0~{\rm for~}n<0~{\rm or~}m<0~;\,c_{0,0}=-c_0\,. 
\eeq

Actually it is not necessary to  explicitly solve these recursion relations: 
it is sufficient to calculate  the first few coefficients and realise 
 that they coincide with those of the Taylor expansion of
\beq
-c_0\sqrt{1+\Tr h+\um(\tr h)^2-\um\tr (h^2)}=-c_0\sqrt{-g}\,.
\eeq
Combining this simple fact with the observations that the solution of the above recurrence relations is unique, that $\int d^2\xi \sqrt{-g}$  is Lorentz 
invariant since \cite{Gliozzi:2011hj}
\beq
\de\sqrt{-g}=-\ep^{aj}\pa_a (X^j\sqrt{-g})\,,
\label{varg}
\eeq
and that of course any constant $\sigma$ is  Lorentz invariant,
we are led to conclude that the most general invariant of scaling zero is the one quoted in the Introduction in Eq. (\ref{zeros}).

  \section{Scaling two}

At higher scaling graphs with vertices with more than one solid link appear.  
 We call seed graphs those of them without vertices of scaling zero, 
i.e. vertices associated with $\p_a X_i$. 
Even in this case, things do not complicate too much, because of the same two features of the variations. First of all, Lorentz variations which involve non 
derived fields  have the same role as above in forming total derivatives. 
Once more, one should add a ring with two vertices to complete the total 
derivative. This new ring takes 
with it the whole tower of graphs already considered at scaling zero. 
It is straightforward to conclude (and verify) that all variations involving non derived fields are exactly compensated by multiplying every graph 
by the scaling zero Lagrangian density. In other words, a Lorentz invariant of scaling $n>0$ should have the form $\sqrt{-g}\,F_n$, where $F_n=\sum_\alpha 
t^\alpha_n$ is a suitable linear combination of terms of scaling $n$. We have
\beq
\delta t^\alpha_n=-\ep^{aj}\left(X_j\pa_at^\alpha_n+
{\rm terms~involving~only~field~derivatives}\right)\,;
\eeq
combining the first term of this transformation with the way of transforming of
$\sqrt{-g}$ given in (\ref{varg})  we obtain a total derivative;
thus, from now on, we will concentrate on that part of Lorentz variation involving only field derivatives and describe a general method to find the linear 
combination $F_n$ where this Lorentz variation is cancelled, i.e.
\beq
\delta F_n=-\ep^{aj}X_j\pa_aF_n\,.
\label{varf}
\eeq 

The Lorentz variation of a vertex with $n$ derivatives is
\beq
\begin{split}
\de \left(\p_{a_1 a_2\dots a_n}^nX_i\right)
&=-\ep^{bj}\Bigl(\sum_k\p_{a_k}X_j\,
\p_{b \,a_1a_2\dots a_n}^{n}X_i+\p_{b}X_i\,
\p_{a_1a_2\dots a_n}^{n}X_j+\\
&+\sum_k\sum_l\p_{a_ka_l}^2X_j
\p_{b\,a_1a_2\dots a_n}^{n-1}X_i+\dots\Bigl)\\
\end{split}
\label{varn}
\eeq
The variations of vertices in seed graphs can be divided into two 
categories: those which generate a vertex of scaling zero 
(the first two terms of (\ref{varn}))  and the others. 
The latter will in general make different seed graphs communicate, and will lead just to new combinatorial factors, as we shall see later. However, to 
deal with scaling two corrections it is sufficient to consider the former.

\begin{figure}[ht]
\begin{center}
\includegraphics[width=0.8\columnwidth,height=0.2\textheight]{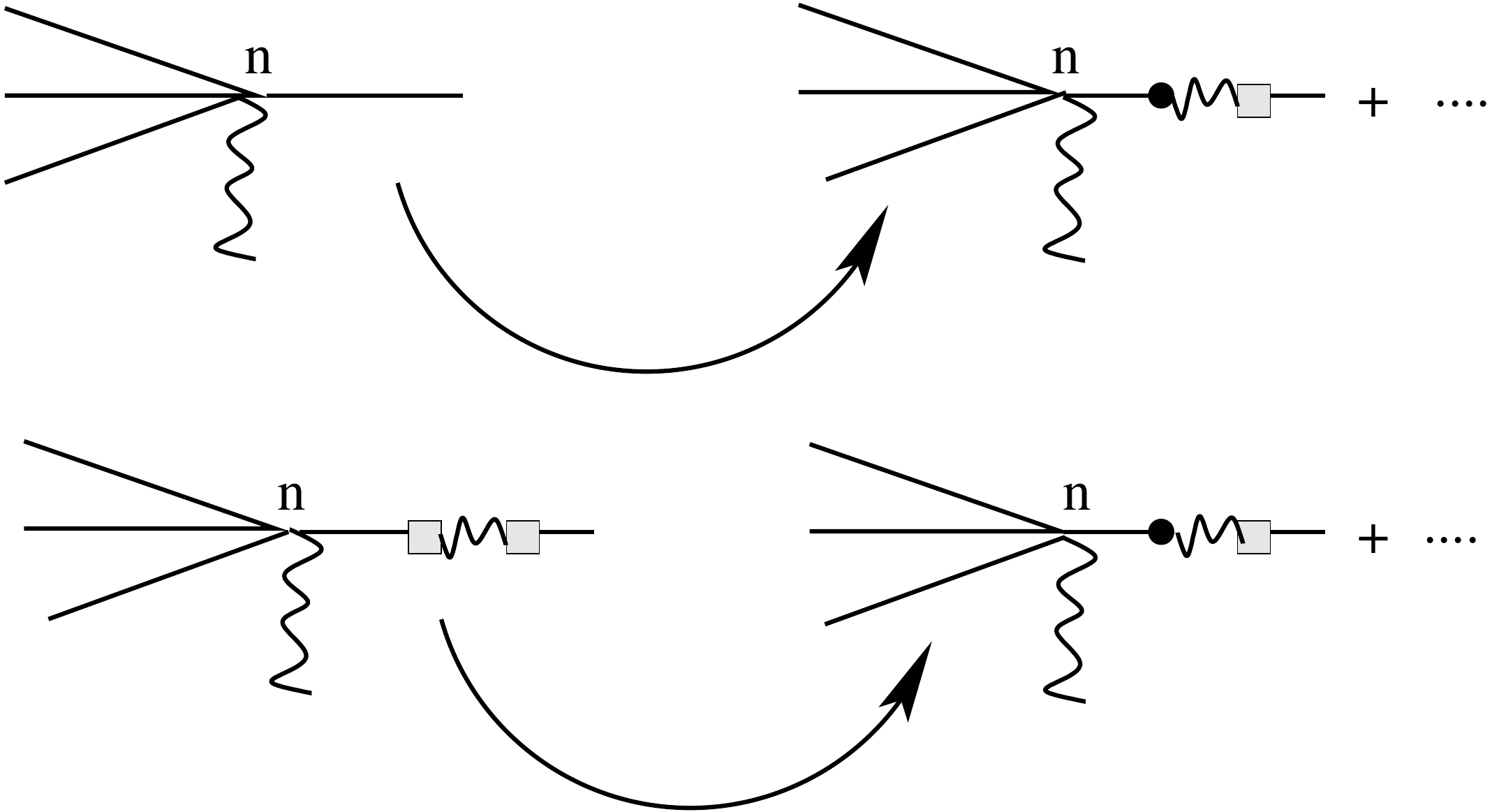}
\end{center}
\caption{In the first line, a variation of the vertex of scaling n. 
In the second line, the variation of the vertex of scaling 0 
on the left, which equals the first one.}
\label{fig:varplain}
\end{figure}

There are two general cases to deal with. These are the first term of (\ref{varn}) whose effect can be disposed of by a modification of the solid link 
stretched between two non-zero vertices, and the second term of (\ref{varn}) 
which requires a modification of a wavy link.  
In fig. \ref{fig:varplain} the  variation induced by the first term is shown. 
Once this first step is made, the chain must grow in order 
to cancel variations which add a dot and a scaling zero vertex. The only 
difference with the ring is that now we can establish an order among 
the vertices. This is sufficient to reduce all multiplicities to one, and to obtain the recurrence relation
\beq
a_{k+1}=-a_k.
\eeq
So every solid link contributes a factor
\beq
\sum_{k=0}^\infty \left[{\big(-h\big)}^k\right]_{ab}=\big(\eta+h\big)^{-1}_{ab}=
g^{ab}\,,
\label{inverse}
\eeq
where $g^{ab}$ is the matrix inverse of the induced metric $g_{ab}$, with 
$g^{ab}g_{bc}=\delta^a_c$. We omitted the arbitrary constant, as it could be absorbed in a unique constant in front of the graph. In conclusion, the variation 
generated by the first term of (\ref{varn}) is cancelled if we replace the 
Minkowski contraction  of every solid link with the induced metric, namely,

\beq
\eta^{ab}\to\,g^{ab}\,.
\label{moveone}
\eeq   

The fact that the infinite tower of graphs generated by (\ref{inverse}) cancels
those Lorentz variations which insert zero scaling vertices 
can also be  verified \emph{a posteriori} by studying the way of 
transforming of $g^{ab}$. Since \cite{Gliozzi:2011hj}
\beq 
\de g_{ab}=-\Lambda_{ab}^{ef}g_{ef}-\ep^{cj}X_j\pa_cg_{ab}\,,
\label{vargab}
\eeq
with
\beq
\Lambda_{ab}^{ef}=\ep^{cj}\left(\p_aX_j\de^e_c\de^f_b+\p_bX_j\de^f_c\de^e_a\right)
\eeq
we obtain
\beq
\de g^{ab}=+\Lambda_{ef}^{ab}g^{ef}-\ep^{cj}X_j\pa_cg^{ab}\,.
\label{varginv}
\eeq
If the seed graph is given by a term of the form
$T_{abcd\dots}\eta^{ab}\eta^{cd}\dots$, it has to be replaced with
$T_{abcd\dots}g^{ab}g^{cd}\dots$. The variation of vertices of $T$ generates
factors of the type $-\Lambda_{ab}^{ef}$, like in (\ref{vargab}), that are 
cancelled by terms of opposite sign generated by the variation of the 
$g^{ab}$'s. Note that the last term of (\ref{varginv}) contributes to the 
total derivative and is taken into account in (\ref{varf}).  
 \begin{figure}[ht]
\begin{center}
\includegraphics[width=0.8\columnwidth,height=0.2\textheight]{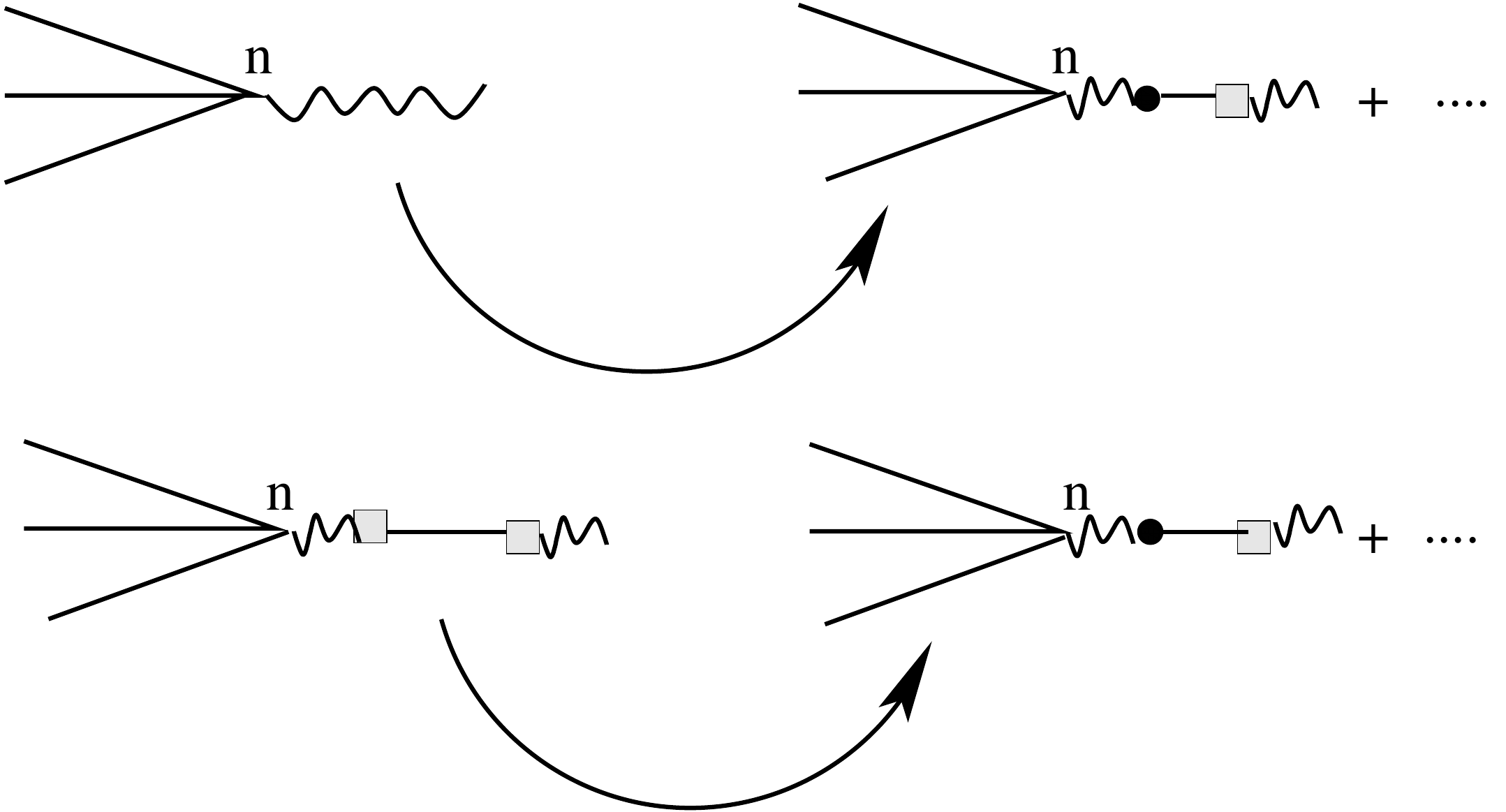}
\end{center}
\caption{In the first line, a variation of the vertex of scaling n. 
In the second line, the variation of the vertex of scaling 0 
on the left, which equals the first one.}
\label{fig:wavy}
\end{figure}

The effect of the second term of (\ref{varn}) on the vertex $X_i$ 
is shown in fig. \ref{fig:wavy}, 
where it is also shown how to cancel this variation by a modification of the 
wavy  line associated with this vertex. The rule is to replace the simple contraction of the wavy line with
\beq
 \de^{ij}\to t^{ij}=\delta^{ij}-\p_aX^ig^{ab}\p_bX^j\,.
\label{movetwo}
\eeq
In this way the mentioned variation is compensated by the variation of $\p_aX^i$.
This move generates a truly new graph only if it is applied to wavy lines 
connecting vertices of non zero scaling. In any other case the second move can be reabsorbed by the first move. For the same reason we cannot iterate this second move more than once for each wavy link.

As already mentioned in the Introduction, if we apply (\ref{moveone}) and 
(\ref{movetwo}) to the two seed terms 
$(\p_{ab}^2X\cdot \p_{cd}^2X)\eta^{ac}\eta^{bd}$ and 
$(\p_{ab}^2X\cdot \p_{cd}^2X)\eta^{ab}\eta^{cd}$ we obtain at once the two 
Lorentz invariants $I_1$ and $I_2$ of scaling two quoted in (\ref{i1}) and 
(\ref{i2}). Again we note that we can  check their invariance immediately 
by  resorting to the transformation law of $g^{ab}$ given in (\ref{varginv}).

 We found no other invariant of scaling two. 
In particular we can prove there is no Lorentz invariant term of 
scaling two of the form $\sqrt{-g}R\,F_0$. For, note that Lorentz invariance, 
without eliminating terms proportional to the equations of motion, 
implies that $F_0$ transforms according to 
 (\ref{varf}) and the only solution with $n=0$ is $F_0={\rm constant}$. 
This may be seen either directly by solving the associated recursion 
relations, or indirectly by noting that if $F_0$ is a solution of 
(\ref{varf}), then  any power 
$(F_0)^k$ is also a solution, hence if $F_0$ were not a constant one would 
generate an infinite sequence of independent Lorentz invariants of scaling two. 
In conclusion, the only Lorentz invariant of the form  $\sqrt{-g}R F_0$ is proportional to $I_1-I_2=\sqrt{-g}R$.
Note that this does not contradict the claim of  \cite{Aharony:2011gb}, 
where it was shown
that a specific term of scaling two has a Lorentz transformation that is
proportional to the equations of motion (we do not allow such a
transformation in our analysis)\footnote{ We benefited 
by an exchange of e-mails with O. Aharony about this point.}.

\begin{figure}[ht]
\begin{center}
\includegraphics[width=0.9\columnwidth,height=0.2\textheight]{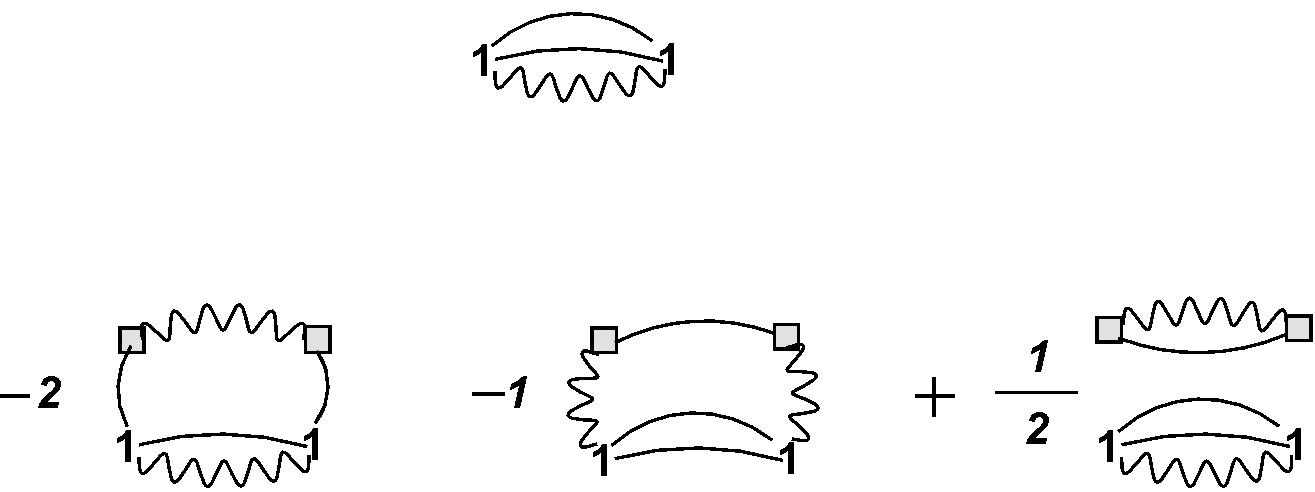}
\end{center}
\caption{In the first line, a seed graph of scaling two. In the second line, the sum of graphs which cancel its variation.}
\label{fig:11}
\end{figure}

What is the first non vanishing contribution of $I_1$ and $I_2$ to the effective action? Consider the seed graph in fig.\ref{fig:11} which generates the  
invariant $I_1$. It is a total derivative, being proportional, after integrating by parts, to the 
free e.o.m.. However, its first correction, according with the rules 
established so far, is a combination of 
six-derivative terms which -- in the case of a $p-$brane with $p>1$ -- does not form  a total derivative. Therefore the combination of graphs drawn in figure 
\ref{fig:11} represents the first bulk correction of the Dirac-Born-Infeld 
action, the multidimensional generalisation of the Nambu-Goto action.
In the  case of the effective string this combination is instead a total 
derivative. This can be understood by observing that $I_1-I_2=\sqrt{-g}R$ 
is a total derivative (at least locally) and that the first terms of $I_2$, up to eight derivatives, are proportional to the free e.o.m, hence are 
vanishing at six-derivative order. On the other hand the only term of scaling four with six derivatives, namely
$\p^3_{abc}X\cdot\p^3_{def}X\eta^{ad}\eta^{be}\eta^{cf}$, is a total derivative modulo free e.o.m., thus we are led to conclude that in the effective string action in any space-time dimensions there are no  six-derivative corrections of the Nambu Goto action.
 
\section{Higher scaling}
The number of Lorentz invariants increases rapidly with the scaling. A first class of invariants is formed by the ring of polynomials of the invariants of lower scaling. In fact, if $F_n$ and $G_m$ are two functions of scaling $n$ and $m$ 
obeying (\ref{varf}), then $(F_n)^p(G_m)^q$ for any pair of integers 
$p$ and $q$ fulfils the same transformation law, so it defines the Lorentz invariant $\sqrt{-g} (F_n)^p(G_m)^q$ of scaling $pn+qm$. Thereby the two invariants of scaling two generate four invariants of scaling four of the form 
$I_\alpha I_\beta/\sqrt{-g}$   $~(\alpha,\beta=1,2)$. A combination of them gives the geometric invariant $\sqrt{-g}R^2$, of course. The first non-vanishing contribution is an eight-derivative term that comes from $I_1^2/\sqrt{-g}$.

In addition to this polynomial class of invariants there are many others 
which can be constructed \emph{ex novo} starting from seed graphs and repeating the steps described in the previous section. For instance, starting from the seed term $(\p_a\p^b X\cdot \p_b\p^cX)(\p_c\p^dX\cdot\p_d\p^aX)$  
and applying the two moves (\ref{moveone}) and (\ref{movetwo}) we get at once 
the Lorentz invariant
\beq
I_3=\sqrt{-g}\,t^{ij}t^{kl}\p_{ab}^2 X_i \p_{cd}^2X_j\p_{ef}^2X_k\p_{gh}^2X_l
\,g^{ha}g^{bc}g^{de}g^{eg}\,,
\eeq  
which is related, when added to a suitable combination of 
$I_\alpha I_\beta/\sqrt{-g}$, with the geometric invariant 
$\sqrt{-g}R_{ab}R^{ab}$, where $R_{ab}$ is the Ricci tensor. In our 
notation we have simply
\beq
R_{ab}=g^{ef}t^{ij}\left(\p_{ae}^2X_i\p_{fb}^2X_j-\p_{ab}^2X_i\p_{ef}^2X_j\right)\,.
\eeq
 In the case of the effective string the equality $R^2=R_{ab}R^{ab}$ holds, 
so $I_3$ is not a new invariant.

 \begin{figure}[ht]
\begin{center}
\includegraphics[width=8.5 cm]{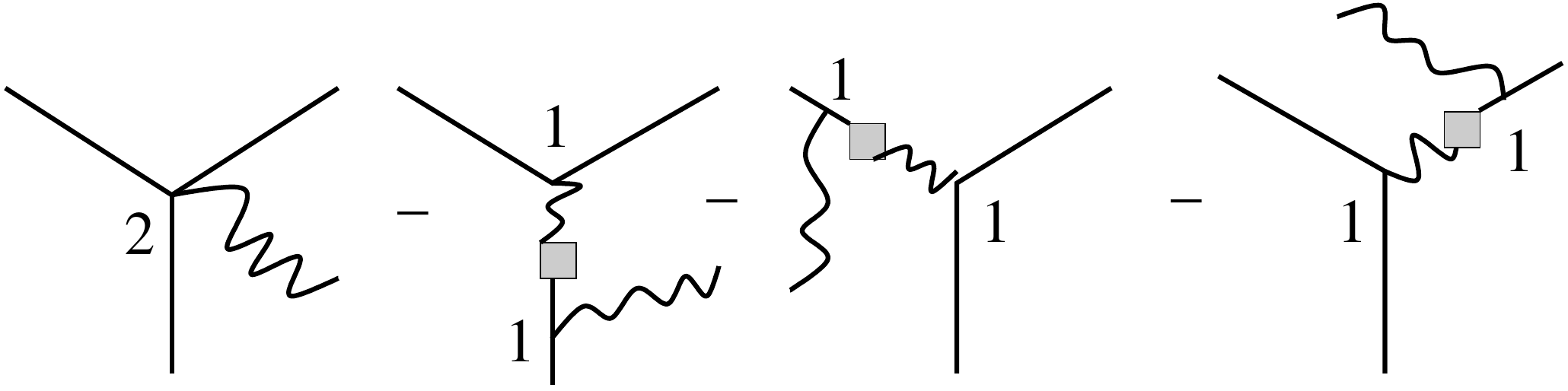}
\end{center}
\caption{Splitting a vertex of scaling two.}
\label{fig:vertex}
\end{figure}

When in the seed graph there are vertices with scaling $n>1$, i.e. transverse coordinates with more than two derivatives, we have to add a third move in 
order to compensate the effect of that part of the variation of $d^nX_i$ which 
creates vertices of scaling $n>0$, like the third term of (\ref{varn}). Analysing the structure of this variation it is easy to see that the rule that 
compensates this kind of variation is to add to each  vertex with $n>1$ 
 all possible splittings of it in a pair of vertices of lower scaling. 
In particular a vertex of scaling two is replaced by
\beq
\p_{abc}^3 X_i\to\nabla_{abc}X_i=\p_{abc}^3 X_i-(\p_{ab}^2X_j\p_{d}X_k
\p_{ec}^2X_ig^{de}\de^{jk} +
{\rm cyclic~ permutations~ of \,} abc)\,.
\label{movethree}
\eeq
This is the third move we need to complete the construction of Lorentz 
invariants.
Its diagrammatic representation is drawn in  figure \ref{fig:vertex}.
The generalisation of the last move to vertices with more legs is straightforward. The crucial point is that the split vertex is a sort of covariant derivative, in the sense that under a non-linear Lorentz transformation its variation is
\beq
\de \left(\nabla_{a_1 a_2a_3}X_i\right)
=-\ep^{bj}\Bigl[\p_bX_i
\p_{a_1a_2a_3 }^3X_j +\left(\p_{a_1}X_j\,\p_{b\,a_2a_3}^3X_i+
{\rm cyclic~ perm.~ of \,} a_1a_2a_3\right)    \Bigl]\,.
\eeq
Comparing it with (\ref{varn}) we notice that the problematic part of the variation has disappeared: the split vertex transforms like vertices of lower scaling, where the steps (\ref{moveone}) and (\ref{movetwo}) suffice to build up an invariant expression.  

\begin{figure}[ht]
\begin{center}
\includegraphics[width=8 cm,angle=-90 ]{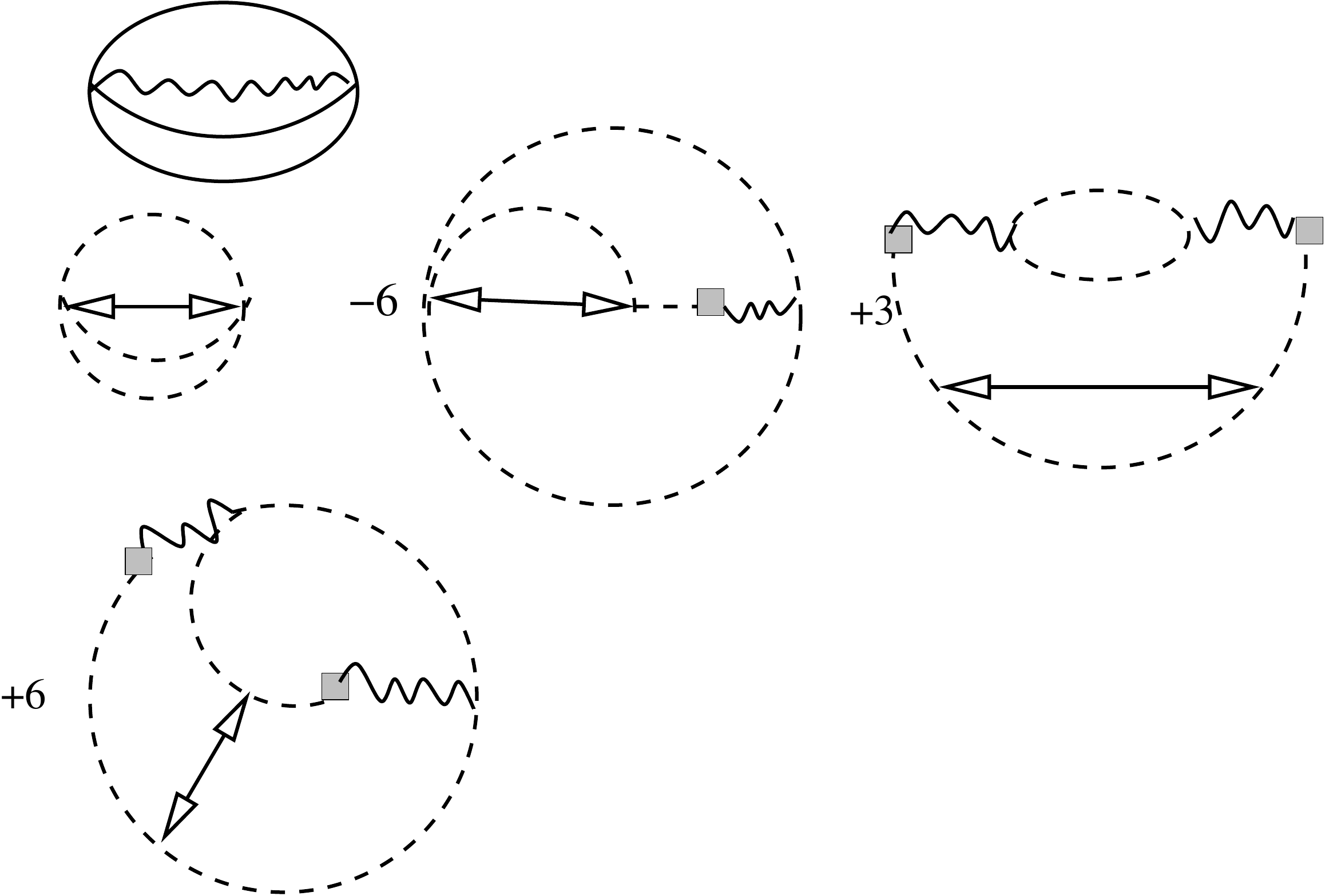}
\end{center}
\caption{In the first line a seed graph with two vertices of scaling two. In the other two lines the Lorentz invariant obtained by applying the moves 
(\ref{moveone}), (\ref{movetwo}) and (\ref{movethree}). The dotted lines represent the saturation with $g^{ab}$ and the lines with arrows represent saturation with 
$t^{ab}$. Note that not all the saturations in the transverse indices (wavy lines) can be promoted to $t^{ab}$, but only those connecting vertices of scaling larger than zero. In order to obtain the complete Lorentz invariant we have to multiply this combination with $\sqrt{-g}$.}  
\label{fig:twovertices}
\end{figure}

If we apply this third move to the seed graph 
$\p_a\p_b\p_cX\cdot \p^a\p^b\p^cX$ we obtain a new invariant of scaling four which is represented in fig. \ref{fig:twovertices}.  
It can be written in a compact way as
\beq
I_4=\sqrt{-g}\nabla_{abc}X_i\nabla_{efg}X_jt^{ij}g^{ae}g^{bf}g^{cg}\,.
\eeq
If one is interested in writing it in an explicit form in terms of transverse coordinates it suffices to use an explicit form of the inverse induced metric. In the case of the effective string we have
\beq
g^{ab}=\frac{h^{ab}-(1+\tr h)\eta^{ab}}{g}\,.
\eeq 

On the contrary the explicit form of $g^{ab}$ is not necessary to 
check directly the Lorentz invariance of the set of expressions one obtains by applying the above three moves. It suffices to know its way of transforming under an infinitesimal Lorentz transformation, described by (\ref{varginv}). This remark suggests a further generalisation of the $p-$brane action. In this context it is customary to assume that among the massless excitations which can propagate in such extended object there is, besides the $D-p-1$ scalars $X_i$, also a $U(1)$ gauge 
field. It has been pointed out that its field strength $F_{ab}$ transforms exactly as $g_{ab}$ under a Lorentz variation \cite{Gliozzi:2011hj}. Thus, if we take the linear combination $e_{ab}=g_{ab}+\lambda F_{ab}$ it follows that its 
matrix inverse $e^{ab}$, with $e^{ac}e_{cb}=e_{bc}e^{ca}=\de^a_b$, transforms exactly as $g^{ab}$. As a consequence, if we replace in our invariants $g^{ab}$ with $e^{ab}$ we obtain more general Lorentz invariants describing the dynamics  of this gauge field. Similarly, in those invariants that can be written only in terms of the induced metric and its inverse, like the geometric invariants, 
we could do the same replacement $g_{ab}\to e_{ab}$ without spoiling the Lorentz invariance of the action. Notice that this way to add higher order terms involving 
$F_{ab}$ was proposed years ago with a different motivation 
\cite{Wyllard:2001ye}.    

\section{Conclusion}
In this paper we described a simple and general method to explicitly construct
higher order Lorentz invariant expressions contributing to the effective action which describes the dynamic behaviour of effective strings or $p-$branes.
We do not know whether the list of invariants constructed this way is 
complete, however the method is so general and the resulting invariant forms 
are so simple that  it would be very surprising 
the discovery of invariants with a different structure.

Summarising the results of the last two sections, we found three simple rules 
which transform a seed term - a term  invariant with respect the stability group $ISO(D-p-1)\times ISO(1,p)$ made with derivatives of the 
transverse coordinates of order higher than one - into an expression which is invariant under the whole Poincar\'e group. They consist in replacing the 
Minkowski metric $\eta^{ab}$ of worldvolume indices with $g^{ab}$, the 
Euclidean metric $\de^{ij}$ on the transverse indices with the metric $t^{ij}$ 
defined in (\ref{movetwo}) and  the derivatives of order higher than two 
with  covariant derivatives, defined in the simplest case in (\ref{movethree}).
Once these replacements have been made the Lorentz transformation of the 
field derivatives is exactly compensated by the transformation law of $g^{ab}$.
We found in this way two invariants of scaling two which can be combined to form the Hilbert-Einstein Lagrangian $\sqrt{-g}R$ which is the first non vanishing 
higher derivative contribution of the effective action of a $p-$brane with 
$p>1$, while for the effective string this term is a total derivative. In the latter case the first non vanishing correction of the Nambu-Goto action is the 
term of scaling four $\sqrt{-g}R^2$.

We can associate a Lorentz invariant to every seed graph, i.e. an arbitrary 
graph with an even number of vertices subject to the only condition that the coordination number of each vertex is larger than two. Different choices of wavy links may give different invariants. Thus we are led to conclude
 that the number of Lorentz invariants is much larger than those that 
can be written in terms of local geometric expressions as 
functions of the induced metric and its derivatives. Note however that 
we work in the  static gauge and there is no obvious reason to believe 
that all the Lorentz invariants that can be found with the present method 
could be rewritten in a reparametrization invariant form.

\label{sec:conclusions}

\section{Note added}

As it was pointed out in \cite{Meineri:2013}, all the invariants up to scaling four, apart from a constant, are in fact a gauge-fixed form of local geometric quantities
constructed with the induced metric: in particular, they are traces and covariant derivatives of the second fundamental form. The claim made in the conclusion of this
paper is therefore not justified. In fact, one can argue by a general counting argument that there is a correspondence between the invariants constructed with the
method we developed and those built out of contractions of the objects $\nabla_a$ and  $K_{ab}^i$ (extrinsic curvature)\footnote{an exception appears in the case of 
one transverse coordinate, where a tadpole is allowed, but it is uninteresting in the context of confining strings}. 
Indeed, at any given scaling $n$, there is one more vertex than those appearing at scaling $n-1$, and one more independent covariant object, 
namely the $(n-1)$th covariant derivative of the extrinsic curvature. Very recently it was also shown that, starting with an action which realizes Lorentz 
symmetry nonlinearly, the same ghost degrees of freedom (St\"uckelberg fields) which are necessary to introduce 
diffeomorphism invariance also enforce linear Lorentz invariance \cite{Cooper}.
The approach suggested in \cite{Aharony:2010cx} and pursued in \cite{Aharony:2011gb} and in the present work (see also the subsequent work 
\cite{Aharony:2013ipa}), instead, gives a precise identity to the invariant 
objects appearing in the action. In particular, non-local quantities do not appear classically among the low-lying contributions, and our method does not generate them at all, as long as we can see.

We recently were made aware that the nonlinear realization of the Poincar\'e group was used to construct the Nambu-Goto and p-brane actions already in \cite{Ivanov:1999gy}. We thank E. Ivanov for pointing out this to us.

\vskip 1cm

\noindent {\large {\bf Acknowledgements}}
We would like to thank O. Aharony for a fruitful exchange of correspondence, 
and L.Bianchi, M.Bill\`o,
M. Caselle, P. Di Vecchia, L. Fatibene and R. Pellegrini for useful discussions.
\vskip 0.2cm

\vskip 1cm
\providecommand{\href}[2]{#2}
\bibliographystyle{abe}
\begingroup\raggedright

\endgroup


\end{document}